\documentclass[twocolumn]{aastex631}
\usepackage{amsmath}
\newcommand{\eb}{\begin{equation}}
\newcommand{\ee}{\end{equation}}
\newcommand{\masyr}{mas yr$^{-1}$}

\definecolor{rkka}{RGB}{219,66,32}

\shorttitle{USNO Bright Star Catalog}
\shortauthors{Zacharias et al.}

\begin{document}

\title{USNO Bright Star Catalog, version 1}

\correspondingauthor{Valeri V. Makarov}
\email{valeri.makarov@gmail.com}

\author[0000-0002-4873-0972]{Norbert Zacharias}
\affiliation{U.S. Naval Observatory, 3450 Massachusetts Ave NW, Washington, DC 20392-5420, USA}

\author[0000-0003-2336-7887]{Valeri V. Makarov}
\affiliation{U.S. Naval Observatory, 3450 Massachusetts Ave NW, Washington, DC 20392-5420, USA}

\author[0000-0002-6597-9606]{Charles T. Finch}
\affiliation{U.S. Naval Observatory, 3450 Massachusetts Ave NW, Washington, DC 20392-5420, USA}

\author{Hugh C. Harris}
\affiliation{U.S. Naval Observatory, Flagstaff Station, 10391 W.\ Naval Observatory Road, Flagstaff, AZ 86005-8521, USA}

\author[0000-0002-4603-4834]{Jeffrey A. Munn}
\affiliation{U.S. Naval Observatory, Flagstaff Station, 10391 W.\ Naval Observatory Road, Flagstaff, AZ 86005-8521, USA}

\author[0000-0001-5912-6191]{John P. Subasavage}
\affiliation{The Aerospace Corporation, 2310 E.\ El Segundo Boulevard, El Segundo, CA 90245, USA}

\begin{abstract}
USNO Bright Star Catalog (UBSC) is a new astrometric catalog of 1423 brightest stars covering the entire sky, which is published online. It is nearly complete to $V=3$ mag except for three stellar systems. A combined astrometric solution of the Hipparcos Intermediate Astrometry Data and two dedicated ground-based campaigns in 2013 -- 2020 is the basis for this catalog. The astrometric parameters for each star include position coordinates, parallax, proper motion components, and their covariances on the Hipparcos mean epoch 1991.25. 64 percent of the catalog are flagged as known or suspected double or binary stars. UBSC lists 68 stars missing in Gaia EDR3 and another 114 stars without Gaia parallaxes or proper motions. The formal precision achieved for proper motions is comparable to that of Gaia.
\end{abstract}

\section{Introduction and observational data} \label{intr.sec}
USNO Bright Star Catalog (UBSC) is the main product of a multi-year program aimed at extending and improving the Celestial Reference Frame representation over the entire sky with the optically brightest stars, which are important for navigation and situational awareness. It is based on three main sources of observational data. 

\subsection{UBAD astrometry}
The USNO Bright-Star Astrometric Database \citep[UBAD;][]{mun} provides accurate, current-epoch astrometry for 364 bright northern-hemisphere stars, including all but five such stars with $V < 3.5$ or $I < 3.2$ and $V < 6$.  Observations were obtained
from 2016 -- 2020 using a 2048 $\times$ 4092 e2v CCD on the USNO, Flagstaff Station, Kaj Strand 61-inch Astrometric Reflector. The bright target stars were imaged through a small 12.5-magntiude neutral-density (ND) spot, allowing them to be astrometrically calibrated directly against much fainter Gaia EDR3 reference stars on the same image.  All observations were
taken using either an $i$ or $z$ filter, and mostly within one hour angle of the meridian, so as to minimize the effects of differential chromatic refraction (DCR).
The exposure times vary
between 15 and 1200 s, with a modal value around 250 s.
This analysis uses the 4807 astrometric measurements on the individual frames (Table~2 of \citet{mun}), rather than the UBAD catalog astrometry itself. 

Astrometric calibration of neutral-density (ND) spot observations bears additional complications relative to
regular CCD techniques. Bright stars generate a significant artifact image of the secondary mirror, which is
larger than the ND spot;  the affected part of the detector was not used for calibration. Flat-fielding within the
ND spot area is not possible, so a dome flat field image taken prior to ND installation was used, possibly
degrading the astrometric performance on bright stars. A windowed moment-based centroiding with SExtractor
was used for star images.
Much fainter Gaia EDR3 stars outside the exclusion area were used for plate calibration.
The field calibration was modeled as affine transformations, with additional residual maps determined and applied separately in $i$ and $z$.

Most of the 364 stars in the UBAD catalog target stars that are listed in the Gaia EDR3
catalog \citep{2021A&A...649A...1G}. Measured positions on the individual frames have been compared with Gaia astrometry, after propagating the Gaia positions to the epochs of the UBAD observations using Gaia proper motions and parallaxes. These UBAD$-$Gaia residuals show no systematic trends with magnitude \citep[Figs. 9 and 10 in][]{mun}, but have median offsets in $i$ of 8.5 and -6.6 mas
in right ascension and declination, respectively, and in $z$ of 11.1 and -9.3 mas in right ascension and declination. The offsets are explained as the field-dependent residual calibration
missing in the model for the ND spot area. There is no way to distinguish whether the offset for the target stars beneath the spot is due to a tilted filter or part of the general residual map, however, its size is consistent with the rest of the residual map outside the ND spot. These constant corrections were added to all single-frame positions.
The single-frame astrometric precision of UBAD, estimated via the
rms differences with Gaia, is of order 5 -- 6 mas in each coordinate.  An empirical floor of 3.5 mas was quadratically added to all measurement
errors. This brought the median normalized rms differences close to 1 for stars at all magnitudes. 

\subsection{URAT-bright astrometric program}
\label{uratsu.sec}
Survey observations with the USNO Robotic Astrometric Telescope (URAT) were taken from 2015-10-31 to 2018-06-30
at Cerro Tololo InterAmerican Observatory (CTIO) \citep{zac19}. The survey area covered almost 3/4 of the sky south of
$+25\degr$. A single interference filter was used with a bandpass between 680 and 750 nm \citep{2015RMxAC..46...23Z}. Unlike the UBAD instrument,
an objective diffraction grating was permanently mounted in front of the lens. The central image, which was not used
for astrometry of bright stars, and the first order diffraction images were separated by 42 pixels on the detector. The elongation of the diffraction images across the grating
was not large due to the narrowness of the passband. However, color-dependent systematic effects in the photocenter
position can be expected along the dispersion, which is close to the east-west direction. To mitigate this
effect, astrometric positions were determined from the two 1st order dispersed images located on both
sides of the central image. The chromatic shifts are expected to be approximately symmetric and almost
cancel each other in the average position. The first-order diffraction
provided an estimated 4.7 mag of attenuation.

Four exposures
were taken of each field, of 60, 30, and two 10 s duration. The shortest exposure in combination with grating
attenuation were still not sufficient to observe stars brighter than $R=4$ mag, hence, an additional neutral
density (ND) filter spot was used on one of the CCDs for the brightest targets. The URAT ND spot is directly on the dewar glass and covers about 570 pixels in diameter. This is large enough to fit the first order grating images of a target star.
A possible constant position shift from it or other non-planar surfaces
can not be distinguished from the general residual map.
We also checked
for possible magnitude and color dependent offsets to the limit
imposed by the small sample size. All stars brighter than 4.5 mag
were observed several times a year through the ND spot. A series of fixed-duration exposures were taken
within the same night staggering the pointing by a few arcsec to average down pixel-specific effects.
Several brightest stars still required special 5 s exposures to avoid saturation.

Low-level calibration of fields included subtraction of mean darks determined for the same exposure duration,
and flat-fielding of the entire detector area except the ND spot, where a uniform pixel response was assumed.
The centroiding algorithm utilized the standard UCAC/URAT procedure of 2D circular Gaussian fitting. The fits
yielded the amplitudes, center positions in detector coordinates, local background value, and the width. 
The elongation of the image was separately estimated through image moments analysis. 
It was typically around 1.2 for first-order diffraction images. Single measurement errors (SME)
combined the estimated standard deviation of centroid from the fitted image profile and estimated atmospheric turbulence
terms. The range of estimated SME was 20 to 60 mas mostly depending on the exposure time.

The next level of calibration was to apply field-dependent distortion corrections. The URAT pipeline used
all Gaia DR2 \citep{2018A&A...616A...1G} stars brighter than $R\approx 16$ mag. Each CCD included hundreds to thousands of reference stars.
A separate 2D distortion map for each CCD was generated by binning all the
available position residuals of reference stars. The $68,270$ calibrated two-dimensional measurements were used in the UBSC production.

\subsection{Hipparcos mission astrometry data}
\label{hip.sec}

The Hipparcos astrometric satellite took observations from space in 1989--1993. The main product is a catalog
of astrometric and photometric mean parameters for 118218 stars, mostly brighter than $V=12$ mag. It is an
all-sky survey, which includes all the bright stars of interest for the UBAD-URAT program. The Hipparcos
catalog was published in \citep{esa} along with a number of other products, including the
Double and Multiple Star Annex (DMSA), and the mid-level epoch astrometry collection called Hipparcos Intermediate Astrometry Data (HIAD).

Since the reference epoch of UBSC is 1991.25, no epoch transformations need to be applied to HIAD equations. HIAD contains partial derivatives of
the star abscissa with respect to the five astrometric parameters of the standard
model in equatorial coordinates,
\begin{eqnarray}
d_1 &=\partial a_i / \partial \alpha_* \nonumber\\
d_2 &=\partial a_i / \partial \delta \nonumber\\ 
d_3 &=\partial a_i / \partial \varpi \nonumber\\
d_4 &=\partial a_i / \partial \mu_{\alpha *} \nonumber\\ 
d_5 &=\partial a_i / \partial \mu_{\delta} 
\end{eqnarray}
where $a_i$ is the abscissa in $i$th observation of a given star, $\alpha$
and $\delta$ are the equatorial coordinates, $\alpha_*=\alpha\cos \delta$, $\varpi$ is the parallax, and
$\mu_{\alpha *}=\mu_{\alpha}\cos \delta$ and $ \mu_{\delta}$ are the
orthogonal proper motion components. 

In the small-angle approximation, a linearized equation for the
observed abscissa difference $\Delta a_i=a_{\rm obs} - a_{\rm calc}$,
can be written as
\eb
d_1\Delta x+d_2 \Delta y+d_3 \Delta \varpi+d_4 \Delta \mu_x +d_5 \Delta \mu_y
+ d_6 \frac{\partial a_i}{\partial \epsilon}\Delta \epsilon
= \Delta a_i,
\label{des.eq}
\ee

In this equation, the notations $\alpha *$ and $\delta$ were changed to
$x$ and $y$, respectively,  and the sixth term is
reserved for an additional auxiliary parameter $\epsilon$ which we will
specify later. Eq.~\ref{des.eq} holds
only in the vicinity of a certain point in the 5D parameter space
$\{\alpha, \delta, \varpi, \mu_{\alpha *},\mu_{\delta}\}$,
as long as the corrections to these parameters remain small. They result from
the Taylor expansion to first degree of the basic astrometric equations,
which are intrinsically nonlinear. This linearization may cause systematic errors
in astrometric solutions when the initial guess of the star values is far from
the truth. To avoid this situation, the Hipparcos data analysis consortia iterated the
entire processing and re-generated condition equations replacing the imprecise
input catalog data with intermediate solution values. The right-hand parts are
the ``observed minus calculated" differences between the actual 1D coordinate
measurements and star abscissae computed from the currently assumed star parameters,
instrument calibration models, and satellite attitude. The latter categories of unknowns
are not present explicitly in HIAD equations, because the corresponding fits have been
subtracted from the right-hand parts. Therefore, the available HIAD data represents the
highest level of mission data reductions with, consequently, limited possible applications.
In-depth re-computation of the mission results is not possible with HIAD, because the essential
lower-level instrument data and metadata are not available. For example, the basic angle
separating the two fields of view cannot be recalibrated from the HIAD, because the field
identification and time of observation are not explicitly specified. Still, the available HIAD data are
sufficient for this project at the level of unperturbed, single stars.

The HIAD dataset is not perfect. A number of problems have been noticed, including a fraction of records with
HIPid = 0, of unknown meaning. HIAD also includes stars that are not listed in the catalog at all. The intended
use is to allow recalculation of the five astrometric parameters for stars if some additional constraints
become available, or estimate the degree of perturbation for problematic objects. HIAD has been used
in a number of ad hoc investigations, e.g., for recomputing proper motions of stars with grossly incorrect
initial assumptions \citep{fab}, assessing the quality of Variability-Induced Motion solutions \citep{pou3},
determination of new orbits from the category of stochastic solutions \citep[binarity flag X,][]{gol}. On the other hand, HIAD is
not suitable for alternative re-processing of the entire mission because crucial information is missing.
In the framework of this project, which deals with a sample of brightest stars, the main shortcoming
is related to the fact that the main condition data $d_1,\ldots,d_5$, $\Delta a_i$, and $\sigma_i$ correspond
to a nominal 5-parameter model, whereas for identified binary or double objects, the catalog parameters were
obtained via specialized pipelines with different models and more complex optimization. Therefore, the
initial parameters, which are consistent with the full catalog solutions, are inconsistent with the
astrometric conditions within the same HIAD file.

\section{Preparation of input data}

\subsection{Pre-processing of UBAD data}
\label{prepub.sec}
The input file of UBAD epoch astrometry includes 4807 epoch observations of 364 stars. The range of epochs is MJD  ($\equiv$ JD$-$2400000.5) 57458.1 
through 59142.5. The sample distributions of formal errors of individual RA and DE measurements are strongly peaked around 4.9 mas.
While the
HIAD epoch data can be used as they are, UBAD data have to be transformed to the chosen
reference epoch of 1991.25 = MJD 48348.5625. This concerns the right-hand parts of the condition equations \ref{des.eq}. They are
exactly the ``O-C" differences of position coordinates from UBAD, where the observed position is taken from
the input data file and the calculated position should be consistent with the nominal set of five parameters
from the HIAD file. Each UBAD observation generates two condition equations, one for RA and the other for DE
differences. These equations are constructed in the following way. For each star, the five nominal HIP
parameters $x_H, y_H, \varpi_H, \mu_{xH},\mu_{yH}$ were extracted from HIAD. Using the JPL Horizons online
database, equatorial coordinates of the Sun (solar ephemerides) were tabulated for the MJD interval covering
the duration of UBAD and URAT campaigns. The table was converted into interpolation function using Interpolation
Order = 1. For each UBAD observation, the current position unit vector of the Sun, $\boldsymbol{r}_{\sun}=
(\cos\alpha_{\sun} \cos \delta_{\sun}, \sin\alpha_{\sun} \cos \delta_{\sun}, \sin \delta_{\sun})^T$, is
calculated. Similarly, we compute the epoch position unit vector $\boldsymbol{r}_{\rm obs}$ using the
measured $\alpha$ and $\delta$. The parallactic offset vector is the tangential component of the difference
in positions of the Sun and the target:
\eb
\label{varp.eq}
\boldsymbol{\tau}_{\varpi}=\boldsymbol{r}_{\sun}-(\boldsymbol{r}_{\sun}\cdot\boldsymbol{r}_{\rm obs})\;\boldsymbol{r}_{\rm obs}.
\ee
The equatorial tangential components of the parallactic displacement are the projections of this vector onto
the basis vectors of the local triad, $\tau_{x\varpi}=(\boldsymbol{\tau}_{\varpi}\cdot \boldsymbol{\tau}_x)$, 
$\tau_{y\varpi}=(\boldsymbol{\tau}_{\varpi}\cdot \boldsymbol{\tau}_y)$, where the local east and north direction vectors
are computed in the regular way:
\begin{eqnarray}
\label{tau.eq}
\boldsymbol{\tau}_x &=& \left( \mathbf{(0,0,1)}
\times \boldsymbol{r}_{\rm obs}\right)/\| \mathbf{(0,0,1)}\times \boldsymbol{r}_{\rm obs}
\| \nonumber \\
\boldsymbol{\tau}_y &=& \boldsymbol{r}_{\rm obs} \times \boldsymbol{\tau}_x 
\end{eqnarray}
This procedure is a step away from the rigorous prescription of vectorial astrometry where $\boldsymbol{r}_H$ should
be used instead of $\boldsymbol{r}_{\rm obs}$. However, the introduced error is insignificant in comparison with the
observational precision, and this algorithm was accepted for the sake of practical convenience.

Stellar aberration and refraction have been taken out from UBAD epoch measurements by the data reduction pipeline. When the right-hand parts of differential condition equations
are constructed, the nominal Hipparcos parallax and proper motion have to be subtracted from the observed position
at the epoch of observation. The RA difference is then
\eb
\Delta\alpha* = (\alpha_{\rm obs}-\alpha_H)\;\cos \delta_H-\mu_{\alpha * H}\frac{({\rm MJD}-48348.5625)}{365.25-
\varpi_H\;\tau_{x\varpi}}.
\ee
The right-hand parts thus computed are consistent with the abscissa residuals $\Delta a_i$ given in the HIAD
equations. The required derivatives for the RA and DE conditions are
also computed from the available data.
Finally, the weights used in the final astrometric solution are
equivalent to the formal errors of UBAD position components. UBAD weights are numerically close to the Hipparcos weights for 1D abscissae. The impact of
UBAD observations on the combined solution is significant per observation, but there are fewer observations available.

\subsection{Pre-processing of URAT data}
\label{prepur.sec}

A total of 159,000 bright star exposures were taken at CTIO for the URAT program and about 1100 of those were taken under the ND spot. The mean epoch of URAT observations is MJD 57652 = 2016-09-21. The URAT data points were selected by a cone of
$2.5\arcsec$ radius around extrapolated Hipparcos positions.
The number of cross-matched stars is 1211 with 20409 matched observations. 
The next step is to assess the performance of URAT epoch data and determine if observations of bright
stars outside of the ND filter could be used in the production of UBSC. Extrapolated Hipparcos positions
are computed for all matched URAT observations using exact URAT epochs. In computing
the tangential components of Hipparcos parallax displacement on the epoch of URAT measurements,
we can replace the parallax vector $\boldsymbol{\tau}_{\varpi}$ with the scaled solar vector 
$\varpi\;\boldsymbol{r}_{\sun}$, introducing a negligibly small error due to the fact that the
barycentric extrapolated Hipparcos position is used instead of the observed position for the tangential
projection. The resulting O-C residuals reveal that both the dispersion and systematic errors of observations outside the ND filter are much greater than those  taken through
the neutral density filter. The distribution of individual residuals is much stretched in east-west direction, which is not fully captured by the formal errors being all between 40 and 72 mas (median 50 mas).
We recall that the URAT grating disperses light into a short spectrum mostly in the RA direction in observations
taken at small hour angles. The  2D Gaussian
fitting centroids (\S \ref{uratsu.sec}) may not be adequate on dispersed and saturated spectra. Furthermore, URAT positions are systematically offset
to the west of the calculated Hipparcos
positions. 

\begin{figure}[h]
\centering
\includegraphics[width=.48\textwidth]{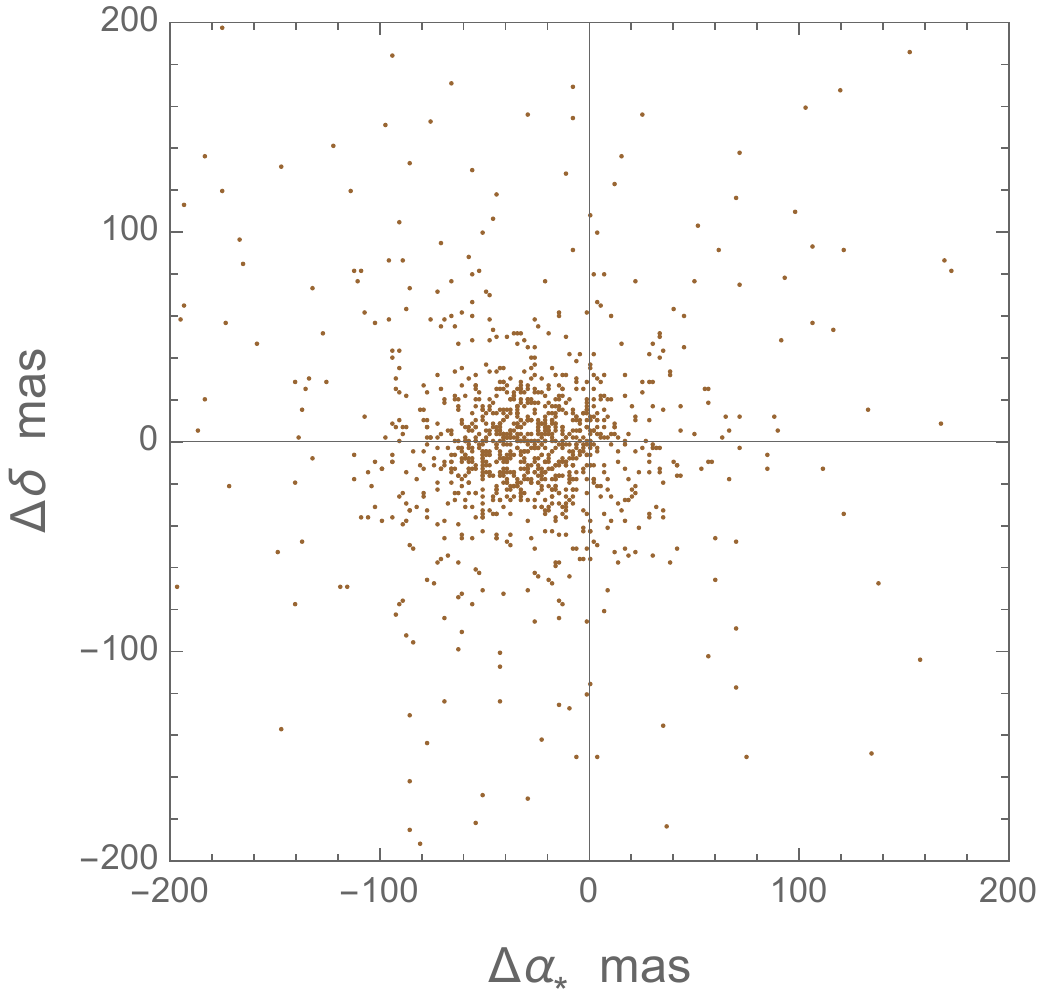}
\caption{Median coordinate differences URAT $-$ Hipparcos for 1211 bright stars. \label{urat3.fig}}
\end{figure}

To determine the systematic corrections to URAT epoch data,
we compute the median RA and DE
position differences  for each star, as well as Median Absolute Deviations (MAD), which
are robust statistics replacing standard deviations. The overall distribution of URAT-HIP median position residuals (with respect to extrapolated Hipparcos positions) is shown
in Fig. \ref{urat3.fig}. The overall scatter in this plot is caused by errors of Hipparcos proper motions and random errors of URAT measurements. We find a clear systematic offset in the RA component. The most likely reason is a bias in the estimated centroids of the spectra convolved with the interference filter transmissivity in the direction of spectral dispersion. The median values of the median coordinate differences are equal to
$-29.4$ mas for RA and $-1.5$ mas for DE. These correction are applied in the combined Hipparcos-UBAD-URAT (HUU) solution, i.e., these numbers were subtracted
from all URAT epoch determinations.

We also check for a possible magnitude equation by plotting the same median residuals versus $H_p$ magnitudes. In the RA component,
apart from the previously mentioned general offset, there is a marginally significant negative (declining) trend with
magnitude between 2 and 6. No corrections for this trend has been applied in the production of UBSC because its origin is not clear. When the coordinate residuals versus the $B-V$ color are investigated, a negative dip in the RA component seems to be present
around $B-V=1.5$ mag. This is probably related to the emergence
of the deep TiO bands at the blue end of the interference filter passband for M-type giants and supergiants, which are
represented in the sample in significant numbers. This feature pushes the photocenter toward the red end of the
spectrum, which is the west direction on the sky for observations near the local meridian. We have not attempted to correct the data for this feature.
There is no certainty that some of the effects seen in the epoch data belong to URAT. For example,
very red stars are also problematic for Hipparcos \citep{pla}. The problem there came from incorrectly estimated
$V-I$ colors, which were also used for color-dependent calibration of the main instrument.

The color-dependent terms of atmospheric refraction are calculated and applied
to the epoch URAT data. We make use of the model of refraction described by \citet{sto}. The monochromatic
refraction is approximated as a series in powers of $\tan{z_0}$, where $z_o$ is the zenith distance, with coefficients that are functions of the air temperature, the index of monochromatic refraction, and the ratio of the local gravity to the sea level gravity. The refraction parameter is modeled as in \citet{owe}, which involves the atmospheric pressure and the water vapor pressure assuming a fixed dew point. The differential chromatic refraction depends on the spectral energy distribution of the observed star and a number of transmission functions. Their effect is greatly reduced by the narrowness of the filter transmission curve, which was assumed to be a tophat function spanning the interval 680 and 750 nm. We assumed a blackbody radiation SED for all stars and an effective temperature of 5040 K for the monochromatic refraction applied in the URAT pipeline.  There
is no uniform and comprehensive source of effective temperatures for the brightest stars. Therefore, 
Johnson $B-V$ colors from the Hipparcos catalog were converted to $T_{\rm eff}$ using E. Mamajek's table of stellar
parameters for the main sequence \footnote{\url{http://www.pas.rochester.edu/~emamajek/EEM\_dwarf\_UBVIJHK\_colors\_Teff.txt}}. The 
differential refraction was computed separately for each URAT observation by numerical integration of the transmitted flux. The computed corrections rarely exceed
3 mas either way. 

\section{HIAD-UBAD-URAT (HUU) combined solution}

An early decision was made at the design stage to not use any data from the Gaia mission apart from the low-level calibration of UBAD/URAT plates using large numbers of fainter stars, and to obtain the new observed
 positions from the UBAD and URAT programs on the Gaia coordinate
 system. The brightest
stars observed by Gaia with different gating methods have distinctly different systematic errors compared to the bulk
of fainter ``regular" targets \citep{lin}. The formal errors of Gaia mean positions, on the other hand, are much smaller than those of UBAD/URAT observations. Since Gaia-HIP proper motion catalogs and their analyses had been produced by other authors  \citep{bra, ker, 2021ApJS..254...42B}, our main motivation was to make a catalog as independent from Gaia as possible for verification purpose.

\subsection{Elements of a global astrometric solution}
\label{cir.sec}
A major weakness of Hipparcos as a first-epoch astrometric catalog is its limited and complicated link to the International Celestial Reference Frame (ICRF) as defined by a set of distant radio-loud quasars \citep{kov}. Gaia, on the other hand, is much better anchored to ICRF owing to its ability to observe and measure optically faint quasars \citep{mig}. This brings up the possibility to improve the quality of Hipparcos data in terms of systematic errors.
Specifically, the global orientation and spin of the Hipparcos frame could be investigated. As a pilot study, we perform a limited global solution at the level of HIAD to demonstrate the technical possibility of correcting systematic errors in Hipparcos.

Following the exposition of \citep{vsh}, we consider a linearized observational equation of space astrometry which includes three main types of unknowns, namely,
the astrometric parameters, the calibration
parameters, and the attitude parameters. The right-hand side of this equation
is computed as the difference between the observed and calculated abscissa
on fixed reference great circles as explained in \S\ref{hip.sec}. A truly global solution of the problem
would require introduction of hard constraints into this system to remove the inherent 6-rank degeneracy
and a one-step least-squares adjustment of the entire system \citep{ber}. This is not possible with the published HIAD, but {\it a posteriori} systematic corrections of any type can be made directly in the right-hand parts by adjusting the calculated abscissae. If these values are changed, however, the HIAD equations are no longer self-consistent because they implicitly include the other types of unknowns that are shared by other stars. Large-scale sky-dependent corrections emerge as the simplest kind that can still be applied because they only concern the attitude part of the system. This can be implemented by introducing a technical abscissa zero-point parameter for each reference great circle (called orbit in HIAD). The abscissa zero-point correction becomes the 6th unknown $\epsilon$ in Eq. \ref{des.eq}, with the
partial derivative $d_6=1$.

Following the technique described by \citet{ber} in Appendix B,  we perform the QR factorization of the astrometric
part and eliminate it. The remaining blocks for the unknown $\epsilon$ are transformed to 
pseudo-inverse forms and the normal matrix is collected by additive accumulation
of the individual blocks. Indexing requires special care here, because the dense blocks of normal equations
for each star have to be correctly split and placed in the global system of normal equations. The transformed and separated condition equations 
are solved by the regular weighted least-squares method yielding a $2341\times 2341$ covariance matrix and
a vector of zero-point corrections. The resulting distribution of 2341 zero-point correction values is nicely centered on zero and the numbers are small with the exception
of a few zero-point extending beyond 0.1 mas. The consistency of the result with the nominal Hipparcos data validates the
HUU algorithm. Even though this adjustment does not produce any substantial changes in the HIAD part of our
solution, this option should be utilized if the opportunity presents itself
to systematically improve the Hipparcos data with the future Gaia data releases.

\subsection{HUU solution for each star}

The final step of HUU processing is to solve the astrometric equations collected from HIAD and
pre-processed UBAD/URAT data. The previously computed zero-point corrections are subtracted from the
right-hand parts using the orbit identification number and a specially generated cross-reference table.
The UBAD/URAT conditions had to be transformed to the solution reference epoch MJD 48348.5625 using HIAD
proper motions and parallaxes and the tangential parallax vectors $\boldsymbol{\tau}_{\varpi}$
per Eq. \ref{varp.eq}. The method of solution is weighted least-squares without any filtering of individual
observations. The formal errors of abscissae and epoch positions were used as weights. The emerging solution
5-vector includes the corrections to the nominal HIAD astrometric parameters for each star. The $5\times5$ covariance matrix
is used to compute the formal errors of the astrometric parameters and the 10 covariances published in
the output catalog. All the available equations from HIAD, UBAD and URAT are solved together and once.

\section{Internal verification of HUU solutions}
\label{int.sec}
Since HUU is based on three independent sources of data, internal checks are possible by combining these
sources to estimate the relative performance and consistency. For this purpose, we obtain four separate astrometric
solutions: 1) Combined solution COSOL, using all three sources for 1423 stars, and the basis for the final UBSC catalog; 2)  UBAD-only solution UBSOL, using
UBAD and HIAD for 152 stars in common between UBAD and URAT; 3) URAT-only solution URSOL, using
URAT and HIAD for 152 stars in common between UBAD and URAT; 4) HIAD-only solution HISOL, using only HIAD data for 1423 stars.
The UBSOL and URSOL trials are specifically meant to estimate the relative performance of UBAD and URAT
on the stars in common. 

 \begin{table*}[h]
 \centering
 \caption{General statistics of verification HIAD-UBAD-URAT solutions.}
 \vskip 0.4cm
 \label{sta.tab}
 \begin{tabular}{@{}lrrrr@{}}
 \hline
   Statistics     & COSOL  & HISOL & UBSOL & URSOL  \\
 \hline
1.5$\times$MAD[$\Delta \alpha_*$], mas & 0.29 & 0.15 & 0.36 & 0.43\\
1.5$\times$MAD[$\Delta\delta$], mas    & 0.26 & 0.12 & 0.33 & 0.32\\
1.5$\times$MAD[$\Delta\varpi$], mas    & 0.34 & 0.18 & 0.43 & 0.39\\
1.5$\times$MAD[$\Delta \mu_{\alpha_*}$], mas yr$^{-1}$& 1.10 & 0.18 & 1.34 & 1.52\\
1.5$\times$MAD[$\Delta \mu_{\delta}$], mas  yr$^{-1}$ & 0.83 & 0.14 & 1.10 & 0.93\\
\hline
Median error[$\Delta \alpha_*$], mas & 0.51 & 0.52 & 0.59 & 0.59\\
Median error[$\Delta\delta$], mas    & 0.38 & 0.39 & 0.39 & 0.39\\
Median error[$\Delta\varpi$], mas    & 0.61 & 0.63 & 0.64 & 0.66\\
Median error[$\Delta \mu_{\alpha_*}$], mas yr$^{-1}$& 0.21 & 0.59 & 0.06 & 0.22\\
Median error[$\Delta \mu_{\delta}$], mas  yr$^{-1}$ & 0.21 & 0.47 & 0.06 & 0.21\\
 \hline
 \end{tabular}
 \end{table*}

\begin{figure}[h]
\centering
\includegraphics[width=.48\textwidth]{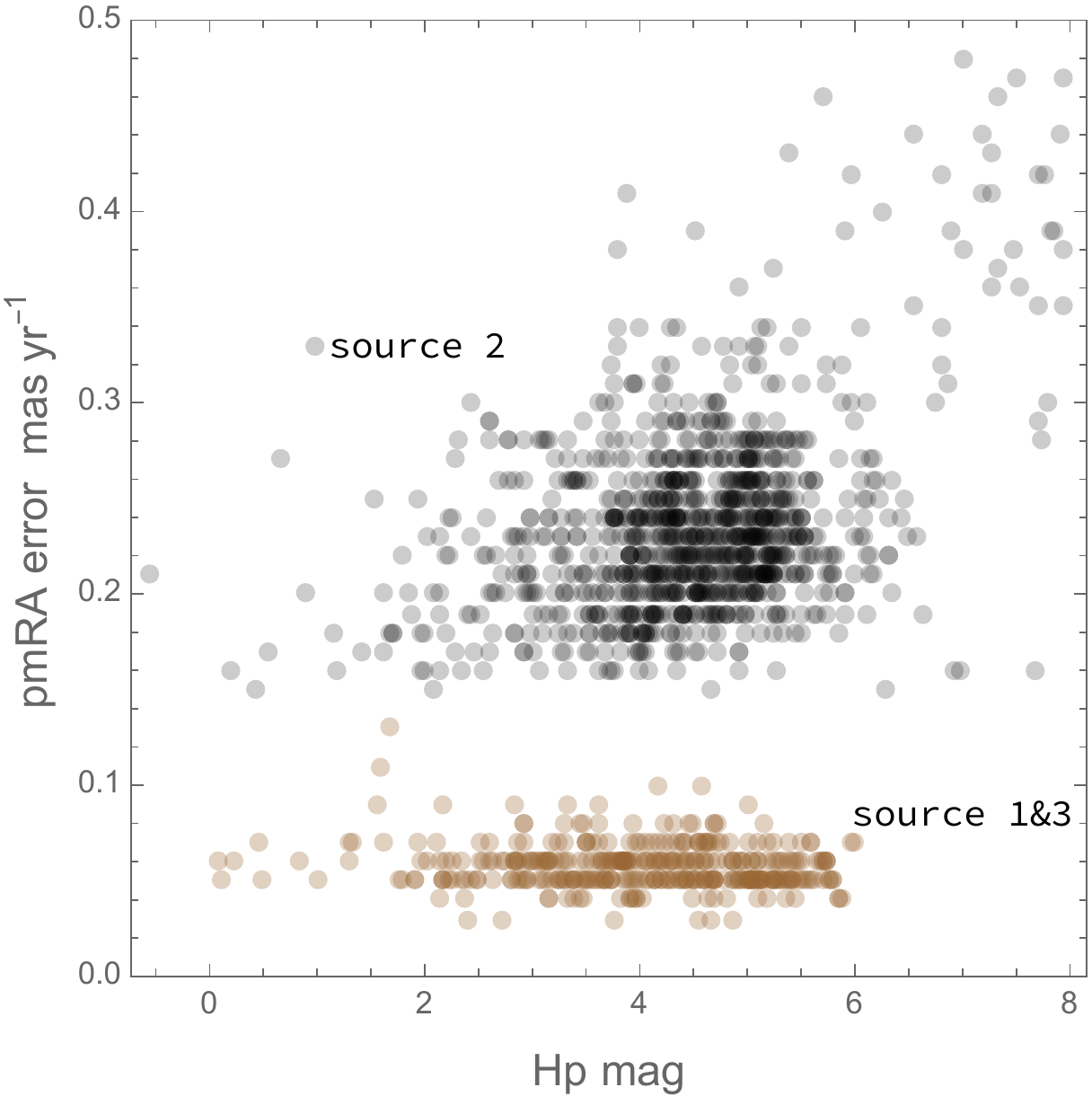}
\caption{Formal errors of proper motion RA components versus $Hp$ magnitude. Source 2 (URSOL) solutions are marked with black circles, sources 1 and 3 (UBSOL and mixed UBSOL+URSOL) with brown circles. \label{pmra.fig}}
\end{figure}

Table \ref{sta.tab} summarizes the overall statistics of these verification solutions. For each of the five
astrometric parameters (which are updates to the nominal HIAD values at 1991.25), the median formal errors
are listed as well as a robust alternative to the RMS residual, namely, $1.5\times$MAD sample value.
Numerically, this construct is close to the standard deviation for a Gaussian sample, but has the advantage
of little sensitivity to the presence of extreme outliers. We will see in the following that a significant
fraction of HUU objects are strongly perturbed, distorting the traditional, non-robust statistical parameters
such as the standard deviation. These deviants are not random flukes or processing artifacts, but most likely
manifestations of astrometric binarity. The largest median formal errors are seen for parallax, as expected,
in the range 0.6 -- 0.7 mas. The formal errors of position coordinates are dominated by the HIAD component because
of the chosen reference epoch, and, therefore, the right ascension is somewhat less precise than declination.
The formal errors of proper motions are appreciably smaller in UBSOL than in the other solutions at 0.06 mas  yr$^{-1}$,
which rivals or possibly surpasses the precision of Gaia proper motions. The reason is the small epoch measurement
formal error of UBAD, which determines the precision of these long-term proper motions. The proper motion errors
in COSOL and URSOL at $\sim0.2$ mas  yr$^{-1}$ are 2.5 times larger than in UBSOL, but still much better than the Hipparcos-only data. A scatter plot of proper motion errors in RA component as a function of $Hp$ magnitude is shown in Fig. \ref{pmra.fig}, where the UBSOL results are shown in brown and URSOL in black.

A more important insight is gained from the statistics of residuals. The scatter of HISOL parameters is much smaller
then all others, because that solution is based on the same data as the nominal solution. The residuals are nonzero
in this case only because of a different weighting scheme and recomputed great circle zero-points. The residuals
of position and parallax are significantly smaller than the median formal errors. The $1.5\times$MAD residuals of proper motion, on the other hand,
are around 1 mas  yr$^{-1}$ in COSOL, which is 5 times their formal errors. This is interpreted as another sign of
wide-spread astrometric binarity.

\section{Catalog contents and format description}
\label{form.sec}

UBSC is almost all-sky complete with the brightest naked-eye stars on the sky with 43 stars present at $H_p<2$ mag
out of 46 in the Hipparcos catalog. Four brightest components of three stellar systems at $H_p<2$ mag are missing, viz., Sirius (too bright for the URAT instrument), $\alpha$ Cen A and B (too bright and too close for URAT), and Polaris (in the north cap exclusion zone for UBAD). The completeness declines with magnitude, with 161 out of 165 at $H_p<3$,
453 out of 481 at $H_p<4$, $1\,003$ out of $1\,473$  at $H_p<5$, $1\,338$  out of $4\,562$  at $H_p<6$ mag due to observational
 limits of R = 3 for UBAD and R = 4.5 to URAT.
The distribution of magnitudes is peaked at $\sim 4.5$ mag, while 85 stars have fainter magnitudes between
6 and 8. A bimodal distribution of $B-V$ colors with two almost detached bumps at $\sim 0$ and 1.5 mag
is characteristic of bright, magnitude-limited samples, revealing the dominance of massive early-type
dwarfs and luminous red giants and a dearth of solar-type dwarfs.

The formal errors of UBSC astrometric parameters have median values of 0.51 mas for RA, 0.38 mas for DE, 0.61 mas for parallax, 0.21 \masyr\ for
RA proper motion, and 0.21 \masyr\ for DE proper motion. These numbers are not adjusted for the reduced $\chi^2$
statistics of residuals. The formal errors of position coordinates and parallax
represent only a modest improvement compared with Hipparcos. The proper motion results, on the other hand, are formally more
precise in UBSC than in Hipparcos by factors 2.3 -- 2.8 because of the much longer time coverage up to 28 years.

The catalog contains records for 1423 stars.
All astrometric parameters and covariances are on the ICRS, epoch J1991.25.
The indexing of astrometric parameters in covariances is : 1) RA, 2) DE, 3) parallax, 4) pm\_RA, 5) pm\_DE. The contents and format are specified in Table \ref{form.tab}.

 \begin{table*}[h]
 \centering
 \caption{Contents and format of UBSC.}
 \vskip 0.4cm
 \label{form.tab}
 \begin{tabular}{@{}lrr@{}}
 \hline
   Column     & unit  & description  \\
 \hline
1 &   &  HIP id\\
2 & mag & Hp magnitude \\
3 & mag & B-V color (from HIP)\\
4 &     & bin\_flag\\
5 &     & source\_flag\\
6 &     & number of observational equations n\_obs in HUU \\
7 &     & $\chi^2$ of residuals\\
8 & mas & MAD of residuals times 1.5 (MAD*1.5)\\
9 & deg & RA\\
10 & deg & DE\\
11 & mas & Parallax\\
12 & mas yr$^{-1}$ & Proper motion pm\_RA*\\
13 & mas yr$^{-1}$ & Proper motion pm\_DE\\
14 & mas & Error e\_RA\\
15 & mas & Error e\_DE\\
16 & mas & Error e\_PAR\\
17 & \masyr\ & Error e\_pmRA*\\
18 & \masyr\ & Error e\_pmDE\\
19 & mas$^2$ & Covariance c12\\
20 & mas$^2$ & Covariance c13\\
21 & mas$^2$ yr$^{-1}$ & Covariance c14\\
22 & mas$^2$ yr$^{-1}$ & Covariance c15\\
23 & mas$^2$ & Covariance c23\\
24 & mas$^2$ yr$^{-1}$ & Covariance c24 \\
25 & mas$^2$ yr$^{-1}$ & Covariance c25 \\
26 & mas$^2$ yr$^{-1}$ & Covariance c34 \\
27 & mas$^2$ yr$^{-1}$ & Covariance c35 \\
28 & mas$^2$ yr$^{-2}$ & Covariance c45 \\
 \hline
 \end{tabular}
 \end{table*}

Binarity/Perturbed flag {\tt bin\_flag} is set to 0
for stars not known to be perturbed and to integers from 1 through 7 for the following categories of objects.
\begin{itemize}
\item
1 for HIP {\tt Mult\_flag} C: Resolved double or multiple stars. Some of them have separate entries in the main Hipparcos catalog, but the
majority are listed with all components only in the  Hipparcos Double and Multiple Star Annex (DMSA). Separations range from
0.1 arcsec to tens of arcseconds. This is the hardest case of perturbation, because the problem can only be
properly addressed at the lowest level of data (transit parameters). 
\item
2 for HIP {\tt Mult\_flag} X: Stochastic (or failed) solutions. They may include orbiting systems with intermediate orbital periods
that are too short for a linear model with accelerations, but too long for an orbital solution. Also,
presumably numerous occurrences of mixed signals such as orbiting system with a variable component, i.e.,
O and V types together, can end up as failed solutions. 
\item
3 for HIP {\tt Mult\_flag} G: Accelerating stars. When the companion is too faint or too close to be resolved, the observed photocenter
can display apparent acceleration caused by the curved orbital motion. This is especially relevant for long-period
binaries when only a segment of the orbit was covered by Hipparcos observations. The standard solution includes
in these cases two additional apparent acceleration parameters, and, sometimes, two additional jerk parameters.
The number of unknowns can be up to 9. Note that the HUU solution always uses the standard 5-parameter model. 
\item
4 for HIP {\tt Mult\_flag} V: Variability-induced motion. Unresolved doubles with a variable component may display coherent variation of
the photocenter position. The Hipparcos solution for this kind has been generally put to doubt \citep{pou3}. The effect is real,
however, and can strongly perturb the astrometric parameters for certain types of variables, such as Cepheids,
Mira-type, short-period eclipsing stars. 
\item
5 for HIP {\tt Mult\_flag} O: Physical binaries with orbital solutions. These are mostly short period systems, for which the 4.2 years of
mission provided  at least most of the orbit ellipse. The data model in this case includes additional 7 parameters
(Thielle-Innes constants, period, eccentricity, and periapse time) and becomes nonlinear. 
\item
6 for astrometric ($\Delta\mu$) binaries: Candidate unresolved binaries with statistically significant differences between the UBSC and Gaia EDR3 proper motions for  that have not been included in the previous categories 1--5. 
\item
7 for astrometric ($\Delta\mu$) binaries: Remaining candidate binaries from the catalog in \citep{maka} that have not been included in the previous categories 1--6. 
\end{itemize}

The percentage of UBSC stars with binarity flags 1 through 5 inherited from HIAD is 24\%. These categories include the special Hipparcos solutions not supported in HIAD. The overall rate of known and suspected binary/multiple objects is a surprising 64\%. The 557 UBSC stars with {\tt bin\_flag}=6 are selected by a direct comparison of the UBSC proper motions with Gaia EDR3 proper motions. The threshold value of the difference in each coordinate, which is $\chi^2$-distributed with 1 degree of freedom, is $\chi^2(1)=3.84146$, which corresponds to a formal confidence level of 0.95. With this relatively loose criterion, stars with {\tt bin\_flag}=6 should be considered as candidate astrometric binaries. The user is advised to take into consideration other quality parameters and the vectorial $\chi^2(2)$ value to estimate the reliability of this indicator in each particular case.

The {\tt source\_flag} is set to 1 for stars with solutions based on UBAD and HIAD only (212 stars), 2 for stars with solutions based on URAT and HIAD only (1058 stars), and 3 for stars with solutions based on all the three sources of data (153 stars).

\section{Verification by comparison with Gaia EDR3 astrometry}

For an external assessment of astrometric quality, the data for common stars in Gaia EDR3 have to be transferred from the mean epoch of UBSC (J1991.25) onto the reference epoch 2016 and the derived quantities directly compared with the EDR3 data. This forward epoch transformation minimizes the impact of Gaia uncertainties in proper motions. The first step is to compute the mean positions at J2016, which is the reference epoch of Gaia EDR3. No epoch transformation is performed for proper motions and parallaxes assuming that the introduced errors are much smaller than the random errors of the UBSC data. The epoch transformation is performed in vectorial astrometry by computing the 3D unit vectors of position and the local triad of coordinate axes and adding the tangential vector of proper motion multiplied by the epoch difference 24.75 yr. Using these positions, a cone search with a radius of $2\arcsec$ of the entire Gaia EDR3 catalog is done and all the matched entries are stored in a separate working file.

The limited search radius carries the risk of missing some of the Gaia counterparts with highly perturbed proper motions. The minimum perturbation or error that can trigger a miss is $2\arcsec / 24.75 =80.8$ \masyr. A total of 1378 cross-matched EDR3 entries were found. The number of UBSC entries with at least one match, however, is only 1355, thus, $1423-1355=68$ stars do not have a Gaia counterpart. A separate cone search with a radius of $20\arcsec$ and a visual inspection of the 68 missing entries verified that no matching counterpart are present in EDR3. The missing stars are mostly very bright, 1st and 2nd magnitude stars. In a few cases, the much fainter binary companions are present in EDR3 but the bright primaries are truly missing.

Whenever more than one counterparts are found in EDR3 (22 UBSC entries), the closest companion by position is taken. Only seven of the more distant companions have parallaxes determined by Gaia. For five pairs with both parallaxes available, physical binarity is likely. Approximately half of this collection have  nearly equal $G$ magnitudes but only one case is found where the more distant companion is explicitly brighter than the closer one.

The following analysis is focused on UBSC stars with binary flags 0 (no known duplicity) and 6 (new candidate astrometric binaries based on Gaia-HUU comparison), see Section \ref{form.sec}. The intersection with EDR3 counts 1036 such stars. 

 Not all of the 1355 cross-matched EDR3 objects obtained adequate solutions in Gaia. We find 114 stars in this sample with {\tt astrometric\_params\_solved} flags set to 3, which means absence of parallax or proper motion determinations. A further 745 entries, which is 55\% of the sample, have this flag set to 95. These objects have a 6-parameter solution in Gaia EDR3 of somewhat degraded quality with one additional parameter called pseudocolor replacing a spectrophotometrically measured effective wavelength. The high rate of weaker astrometric solutions of bright stars in Gaia is an important point in the following comparative analysis.

\subsection{Comparison of positions at J2016}
The extrapolated UBSC positions are dispersed due to the position uncertainties of UBSC at the reference epoch and the accumulated HUU proper motion error, which is approximately $0.21\cdot 24.75=5.2$ mas. The epoch of the smallest position error is individual for each star and, in general, is somewhat later than 1991.25 because of the relative weights of HIAD and UBAD/URAT measurements. 

\begin{figure*}[h]
\centering
\includegraphics[width=.48\textwidth]{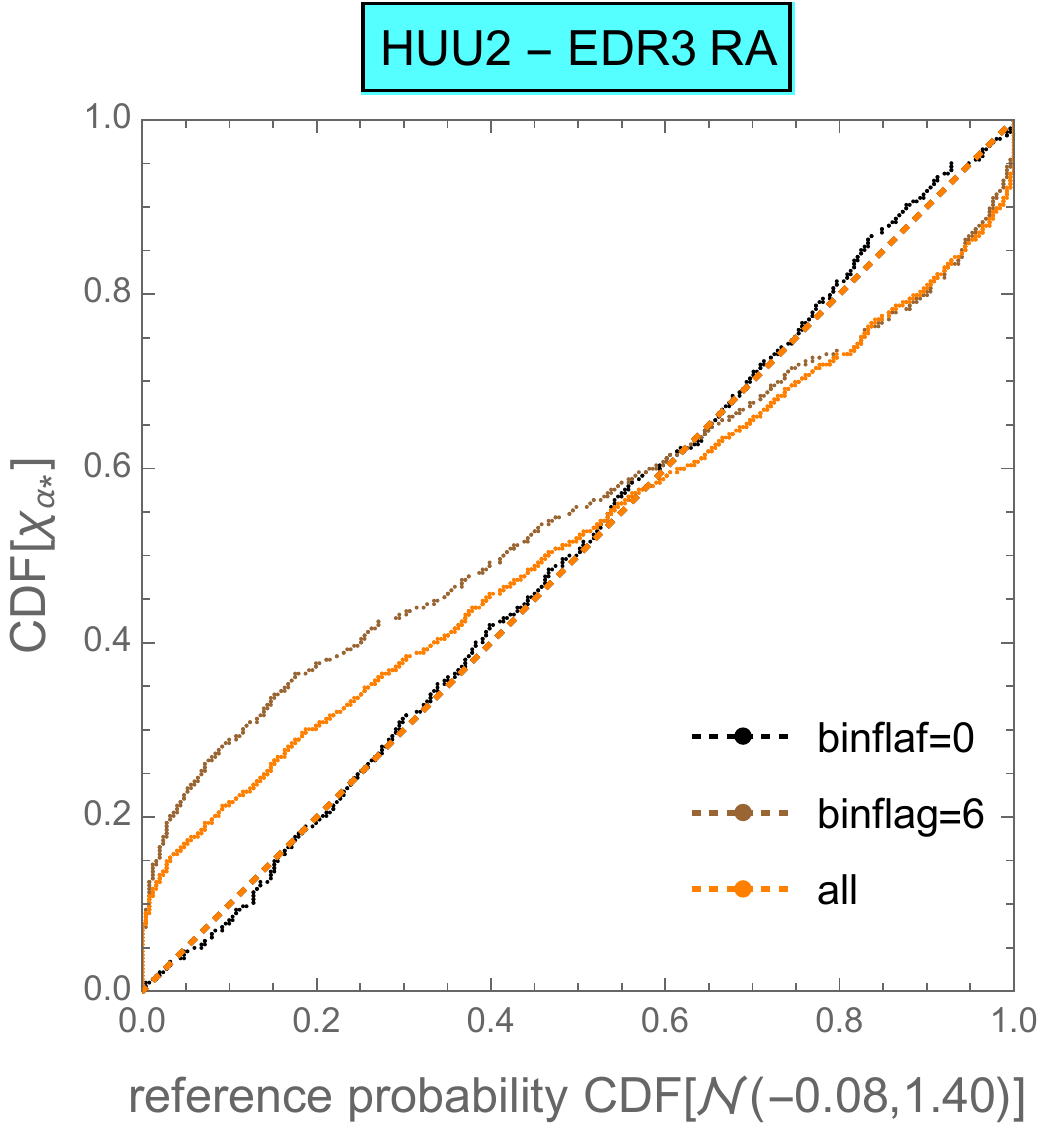}
\includegraphics[width=.48\textwidth]{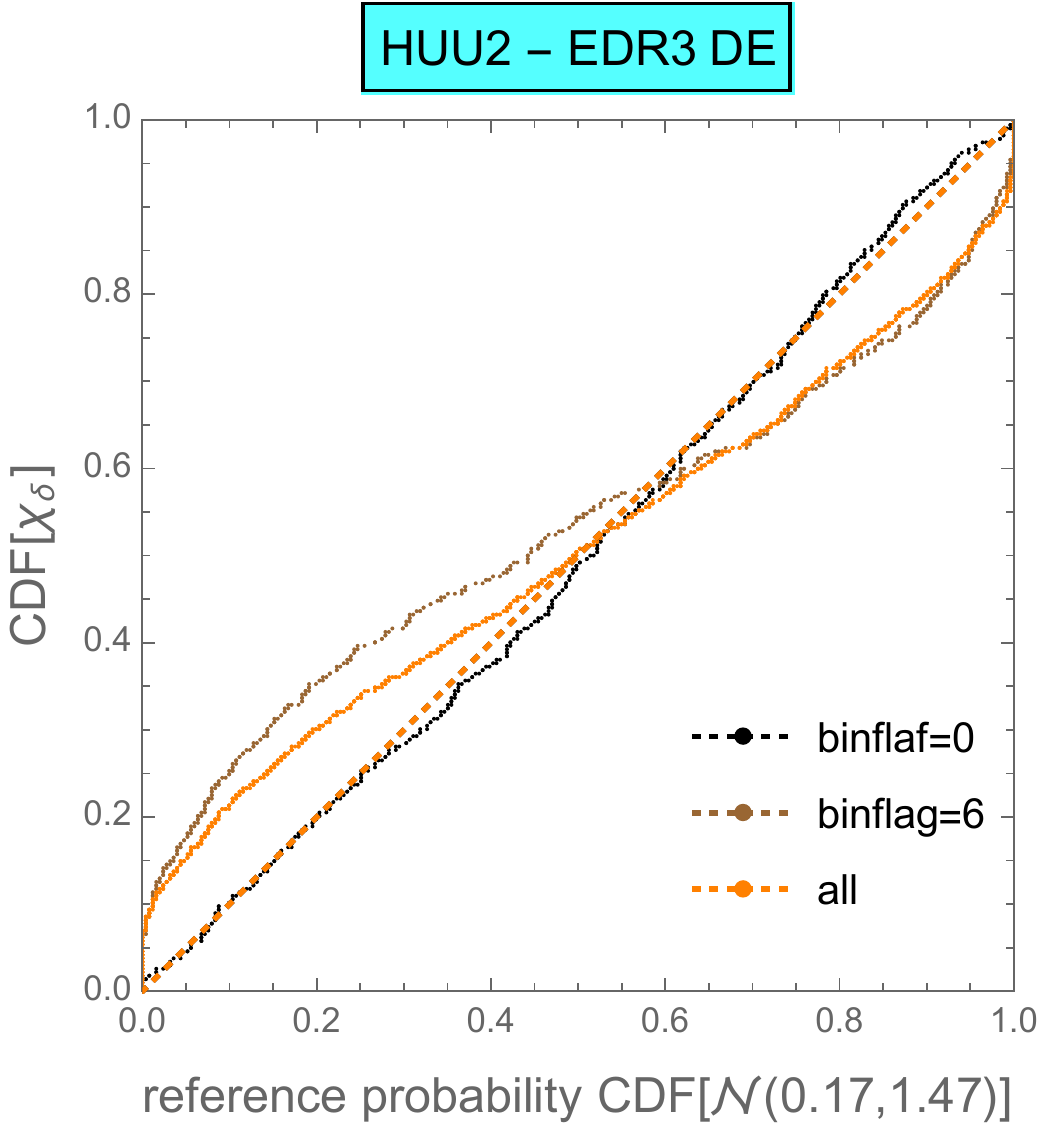}
\caption{Normal probability plots of normalized coordinate differences UBSC(J2016) $-$ EDR3 $\chi_{\alpha*}$ (left) and $\chi_{\delta}$ (right). The diagonal straight line represents the scaling CDF of the {\it fitted} normal distribution for the binflag=0 sample. \label{pos.fig}}
\end{figure*}

If $\boldsymbol r_h$ is the position 3-vector of HUU transferred to the mean epoch of EDR3, and $\boldsymbol r_g$ is the position vector taken directly from the EDR3 catalog, the covariance matrix of the difference $\boldsymbol r_h-\boldsymbol r_g$ is
\eb \begin{split}
\boldsymbol C &= E[\boldsymbol r_h \cdot \boldsymbol r_h^T] + E[\boldsymbol r_g \cdot \boldsymbol r_g^T] \\
&+E[\boldsymbol \mu_h \cdot \boldsymbol \mu_h^T]\,\tau^2+\left(E[\boldsymbol r_h \cdot \boldsymbol \mu_h^T] + E[\boldsymbol \mu_h \cdot \boldsymbol r_h^T]\right)\,\tau,
\end{split}
\ee  
where $\tau=24.75$ is the epoch difference and $E$ is the expectancy operator. All the covariances involved in this computation are available from the UBSC and EDR3 catalogs. The normalized coordinate difference in RA is then
\eb
\chi_{\alpha *}=(\boldsymbol r_h-\boldsymbol r_g)\cdot \boldsymbol e_E/\sqrt{C_{11}},
\label{chic.eq}
\ee 
and similarly for $\chi_{\delta}$. It is also of interest to use the directional part of information in the covariance matrix and compute the 2D $\chi^2_{\boldsymbol r}$ of the tangential position difference vector:
\eb 
\chi^2_{\boldsymbol r}=(\boldsymbol r_h-\boldsymbol r_g)^T\,\boldsymbol C^{-1}\,(\boldsymbol r_h-\boldsymbol r_g).
\label{chi2.eq}
\ee 
This statistic is expected to be distributed as $\chi^2(2)$.

Fig. \ref{pos.fig} shows normal probability plots for the normalized coordinate differences $\chi_{\alpha *}$ and $\chi_{\delta}$ (Eq. \ref{chic.eq}). The straight diagonal lines represent the location of the normalizing probability distribution, which were fitted to the actual sample distributions for stars with binarity flag set to 0. These best-fitting distributions are normal distributions ${\cal N}(-0.081, 1.404)$, ${\cal N} (0.170, 1.469)$, respectively. The black, brown, and orange sequences of dots represent the normalized CDF distributions of the 479 binflag=0, 557 binflag=6, and the entire sample of 1355 stars, respectively. The data points for binflag=0 (single) stars closely follow the diagonal by construction, while those for suspected astrometric binaries and the overall UBSC sample show much greater dispersion of position differences. Even for the currently flagged single stars, the fitted normal distribution suggests a 40--47\% overhead with respect to the formal errors in HUU and Gaia. This can also be seen from the $1.5\times$MAD values, which are a robust statistics proxies for standard deviation, for these three samples collected in Table \ref{1.5mad.tab}.

\begin{deluxetable}{l|lll}
\tabletypesize{\scriptsize}
\tablecaption{Robust statistics $1.5\cdot$MAD for absolute and normalized coordinate differences HUU$-$EDR3 at J2016 \label{1.5mad.tab}}
\tablewidth{0pt}
\tablehead{
\colhead{} & \multicolumn{1}{c}{binflag=0} & \multicolumn{1}{c}{binflag=6} &\multicolumn{1}{c}{all} 
}
\startdata
$\Delta \alpha_*$, mas & 5.76 & 10.35 & 9.50 \\
$\Delta \delta$, mas   & 5.38 & 10.86 & 9.71 \\
$\chi_{\alpha *}$ & 1.41 & 2.57 & 2.14 \\
$\chi_{\delta}$   & 1.44 & 2.57 & 2.27  
\enddata
\end{deluxetable}

The scatter of UBSC$-$EDR3 coordinate differences is approximately 5.5 mas for presumably single stars but almost double that number for newly identified astrometric binaries. By construction, binflag=6 stars have statistically significant proper motion differences between the two catalogs. Most of this additional scatter comes from the proper motion differences extrapolated over 24.75 years. Other potential contributors to the increased scatter are underestimated mean position errors in UBSC and EDR3. There is no sufficient information to analyze the relative weight of these factors. 

\begin{figure}[h]
\centering
\includegraphics[width=.48\textwidth]{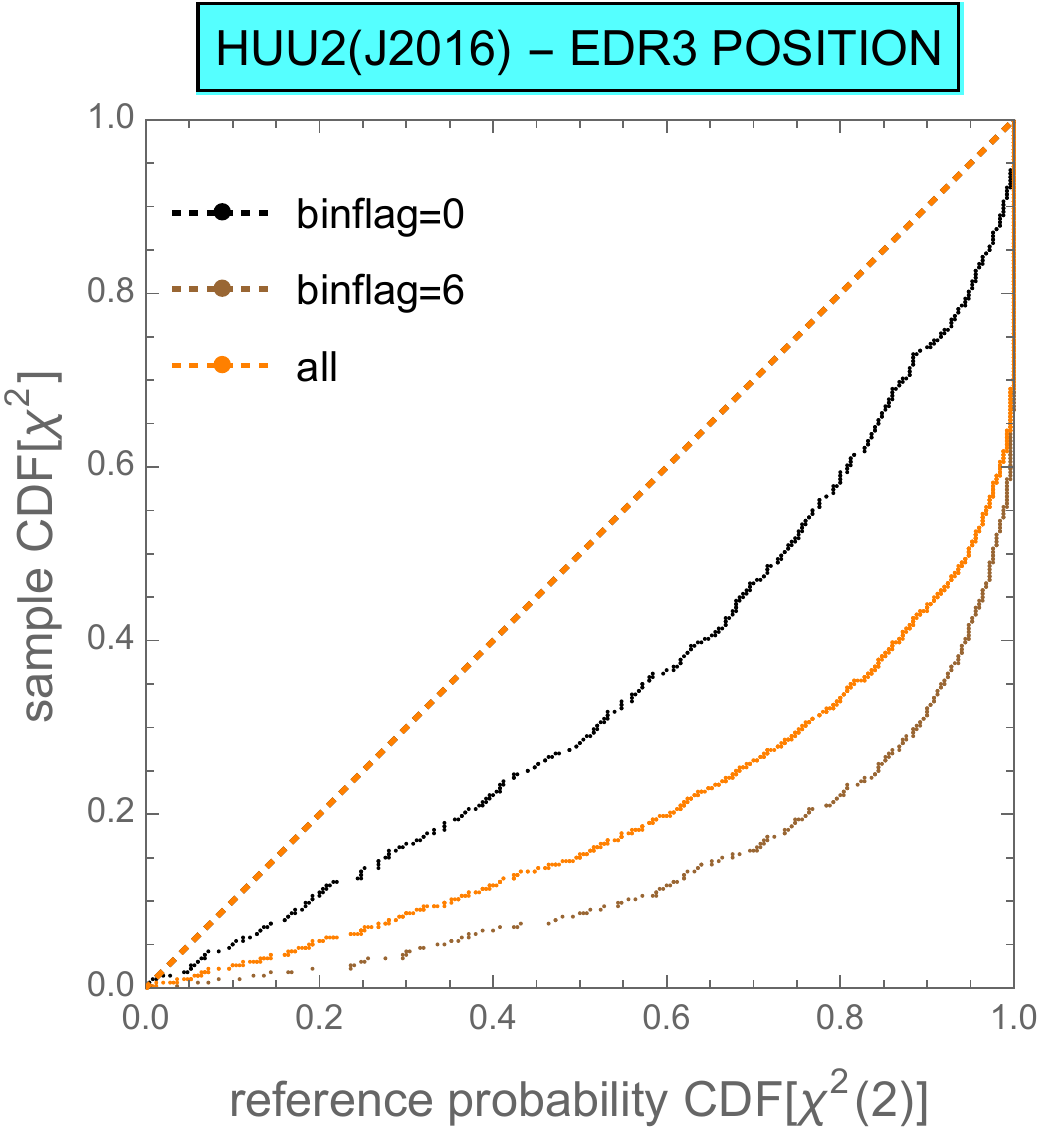}
\caption{Probability plots of normalized position vector differences UBSC(J2016) $-$ EDR3 $\chi^2$. The diagonal straight line represents the scaling CDF of the {\it expected} $\chi^2(2)$ distribution for each sample. \label{r.fig}}
\end{figure}

Preserving the directional information in position difference vectors $\boldsymbol r_h-\boldsymbol r_g$ and their covariances, we also consider the sample distribution of the statistic $\chi^2$, Eq. \ref{chi2.eq}. A normal probability plot for the three samples under consideration is shown in Fig. \ref{r.fig}. In this case, the scaling reference distribution is the theoretically expected CDF$[\chi^2(2)]$, represented by the straight diagonal line. The closest sample distribution to it is that of the binflag=0 sample. Still, it is considerably more dispersed, with a best-fitting distribution of $\chi^2(3.13)$ and a median value of 2.67 instead of the expected 1.39.

\subsection{Parallaxes}

UBSC parallaxes are of limited value because they closely follow the Hipparcos measures. This is the direct consequence of the lower weight of UBAD/URAT measurements in the HUU processing and the suboptimal cadences of ground-based observations for parallax. For the sake of completeness, we compute the HUU$-$EDR3 differences of parallaxes $\varpi_h-\varpi_g$ and the normalized difference $(\varpi_h-\varpi_g)/\sqrt{\sigma^2_{\varpi_h}+\sigma^2_{\varpi_g}}$, where the formal errors of parallax are taken from the UBSC and EDR3 catalogs. There are only 1241 UBSC stars whose Gaia counterparts have parallaxes in EDR3---thus, 114 Gaia counterparts are present with positions but without parallax or proper motion determinations. This makes a total of 182 UBSC stars that do not have parallaxes or proper motions in Gaia EDR3. 

\begin{figure}[h]
\centering
\includegraphics[width=.48\textwidth]{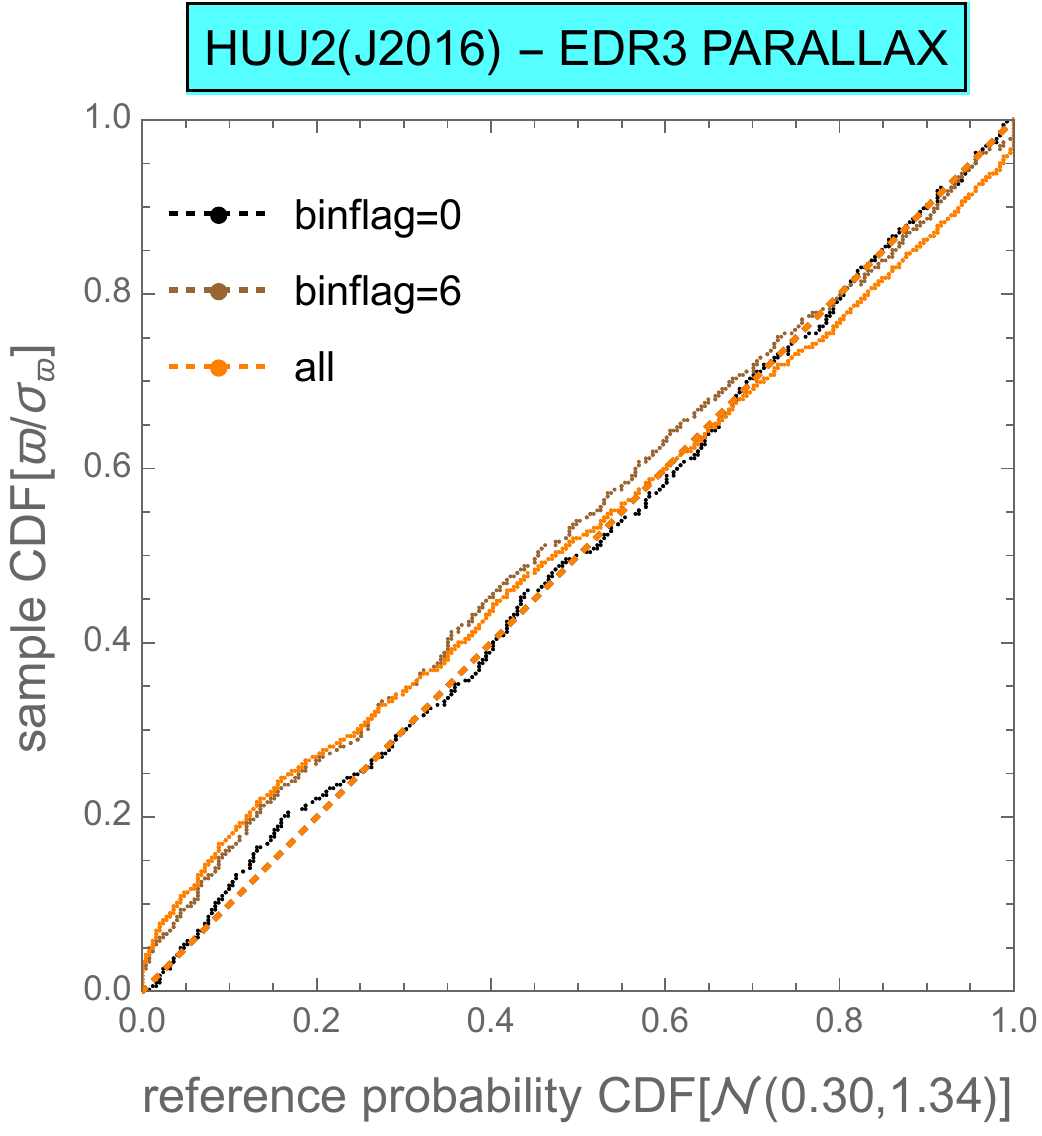}
\caption{Probability plots of normalized parallax differences UBSC $-$ EDR3 $(\varpi_h-\varpi_g)/\sqrt{\sigma^2_{\varpi_h}+\sigma^2_{\varpi_g}}$. The diagonal straight line represents the scaling CDF of the {\it fitted} ${\cal N}(0.304, 1.339)$ distribution for the binflag=0 sample. \label{par.fig}}
\end{figure}

Separating again the samples of binflag=0 (single), binflang=6 (possible astrometric binaries), and all stars, we count 456, 533, and 1241 stars, respectively. Surprisingly, there is a bias of parallaxes for all these samples, with the median parallax of single stars greater in UBSC by 0.17 mas than in EDR3. This may be a problem with Gaia. The bias is confirmed by fitting a Gaussian distribution into the samples of normalized differences, which results in ${\cal N}(0.304, 1.339)$ for binflag=0 and ${\cal N}(0.111, 1.846)$ for binflag=6. Normal probability plots with the former distribution as reference are shown in Fig. \ref{par.fig}. First, we note that the overall dispersion for the samples with binary stars included is not as great for parallax as it is for the extrapolated positions in Fig. \ref{pos.fig}. This confirms that parallaxes are generally more stable to perturbations coming from resolved and accelerating binaries than proper motions. However, there is also a considerable overhead of 34\% in the observed dispersion relative to the expected error. The likely origin of this overhead is HUU because the median formal error of parallax in HUU (0.61 mas) is approximately 4 times the formal error in EDR3 (0.16 mas). The less likely alternative is that the accuracy of EDR3 parallaxes for the brightest stars is catastrophically overestimated. Overall, 80\% of the parallax differences for the binflag=0 sample are within the range $[-0.98,1.21]$ mas.

\subsection{Proper motions}

\begin{figure}[h]
\centering
\includegraphics[width=.48\textwidth]{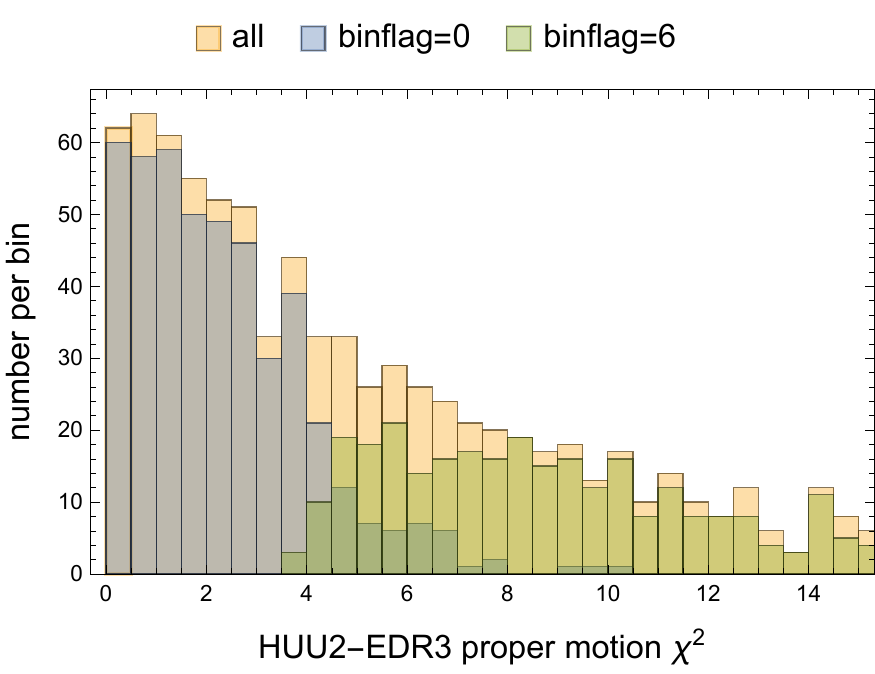}
\caption{Histograms of $\chi^2$ proper motion vector differences UBSC(J2016) $-$ EDR3. \label{pm.fig}}
\end{figure}

The differences in proper motions, $\boldsymbol \mu_h - \boldsymbol \mu_g$, available for 1241 stars, have been used to set the binarity flag 6, signifying suspected astrometric binaries, for 533 stars. The remaining binflag=0 stars only count 456 objects, illustrating the high rate of $\Delta \mu$-binaries among the brightest stars. The other reason for the importance of UBSC proper motions is their assumed accuracy comparable to that of Gaia. Indeed, the median formal error of HUU proper motion for the entire UBSC-Gaia intersection is 0.21 \masyr\ per coordinate, while the median Gaia errors are 0.16 and 0.15 \masyr\ for RA and DE, respectively. 
The distribution of the formal proper motion errors in UBSC is bimodal with the smaller peak around 0.06 \masyr\ representing the HIP+UBAD part of the catalog. Thus, there is no drastic difference in the uncertainty of Gaia and UBSC proper motions. Fig. \ref{pm.fig} shows the distributions of the $\chi^2$ values of proper motion vector differences for the binflag=0, binflag=6, and all common stars with proper motions in Gaia, computed as
\eb 
\chi^2_{\boldsymbol \mu}=(\boldsymbol \mu_h-\boldsymbol \mu_g)^T\,\boldsymbol C^{-1}_{\boldsymbol \mu}\,(\boldsymbol \mu_h-\boldsymbol \mu_g).
\label{chi2mu.eq}
\ee 
The covariance matrix $\boldsymbol C_{\boldsymbol \mu}$ in this case is the sum of the corresponding covariance blocks from UBSC and EDR3. The previously known or suspected Hipparcos binaries show a higher rate at medium $\chi^2$ values between 4 and 7. The overlap between the binflag=0 and binflag=6 distributions (visible as a darker green tone) is caused by the different criterion for identification of candidate astrometric binaries in  UBSC, where proper motion components were used rather than proper motion vectors. 

\begin{figure}[h]
\centering
\includegraphics[width=.48\textwidth]{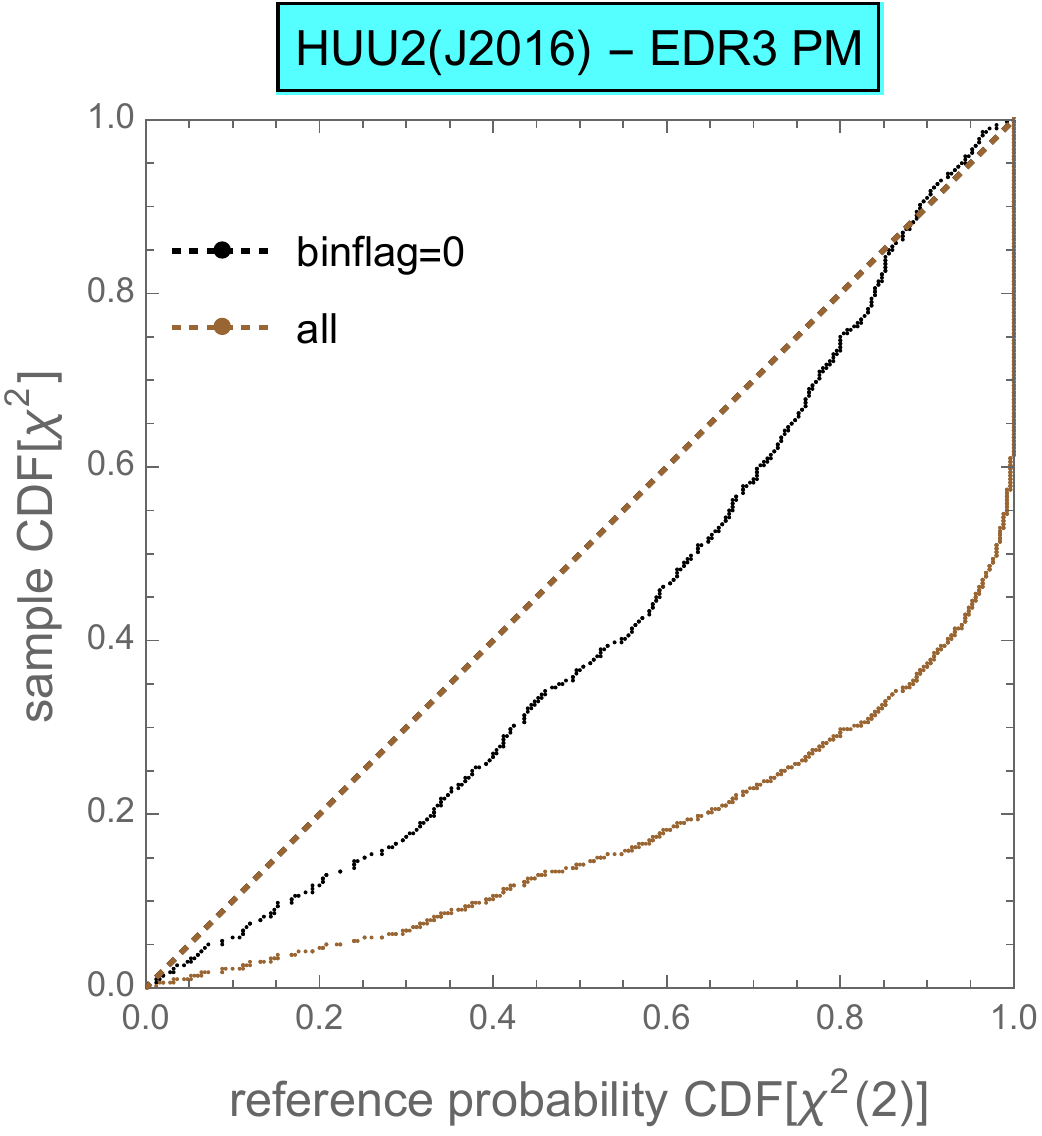}
\caption{Probability plots of normalized proper motion vector differences UBSC(J2016) $-$ EDR3 $\chi^2_{\boldsymbol \mu}$. The diagonal straight line represents the scaling CDF of the {\it expected} $\chi^2(2)$ distribution for each sample. \label{mu.fig}}
\end{figure}

The robust proxy of the standard deviation, 1.5$\times$MAD, for the 456 stars currently identified as single are 0.33 and 0.30 \masyr\ for proper motion differences in RA and DE, respectively. This dispersion is still higher than the expected distribution $\chi^2(2)$ for the 2D differences, as shown in Fig. \ref{mu.fig}. The sample CDF for binflag=0 is sagging below the diagonal for most of the range, but not as drastically as for the general sample, which includes known and suspected binaries. The best-fitting distribution for the binflag=0 sample is $\chi^2(2.44)$.

\section{Conclusions}
Gaia EDR3 does not provide any explicit flags or indicators for double and binary stars. We can therefore only rely on the available information from Hipparcos and the newly established binflag=6, which can be compared with the data from \citep[e.g.,][]{2021ApJS..254...42B} for stars in Gaia EDR3 with measured proper motions. The highest precision in UBSC is achieved for proper motions, which can rival that of Gaia EDR3. The main difficulty in assessing the achieved level of accuracy for the brightest navigation stars is the presence of a large number of visual double stars and unresolved astrometric binaries, which drastically perturb the estimated parameters in both UBSC and Gaia.
On the other hand, proper motions are most sensitive to binarity perturbations. This creates ambiguity of interpretation of UBSC$-$EDR3 differences. These differences are much more dispersed than their formal uncertainties would suggest. The reason may well be the common effect of unresolved $\Delta\mu$ binaries \citep{maka}. Underestimated formal errors in EDR3 cannot be precluded either, although their contribution may only be significant for the proper motion difference but not for the position differences at J2016. The latter suggest that the formal errors of extrapolated UBSC positions are underestimated by 40--45\% even for the stars flagged as single. The likeliest culprit is an underestimated uncertainty of UBAD/URAT position measurements. Alternatively, large systematic errors in Hipparcos could be considered, which currently seem unlikely. UBSC parallaxes are also dispersed by 34\% more than expected (unless Gaia parallax errors for bright stars are catastrophically underestimated). The parallaxes are best aligned between Gaia and UBSC, probably because they are less vulnerable to the perturbations caused by binarity. The utility of UBSC parallaxes, however, is practically limited to the subgroup of star missing in Gaia or included without parallax determinations.

\end{document}